\documentclass[epj]{webofc}
\usepackage[varg]{txfonts}   
\begin{document}
%
%
\title{Present theoretical uncertainties on charm hadroproduction in QCD and prompt neutrino fluxes}
%
%

\author{M.V. Garzelli\inst{1}\fnsep\thanks{\email{maria.vittoria.garzelli@desy.de}} \and
        S. Moch\inst{1}
\and
        G. Sigl\inst{1}
}

\institute{
University of Hamburg, II Institute for Theoretical Physics, D-22761 Hamburg, Germany
          }

\abstract{Prompt neutrino fluxes are basic backgrounds in the search of high-energy neutrinos of astrophysical origin, performed by means of full-size neutrino telescopes located at Earth, under ice or under water. Predictions for these fluxes are provided on the basis of up-to-date theoretical results for charm hadroproduction in perturbative QCD, together with a comprehensive discussion of the various sources of theoretical uncertainty affecting their computation, and a quantitative estimate of each uncertainty contribution.}
\maketitle
\section{Introduction}
\label{intro}

Very large volume neutrino telescopes (VLV$\nu$Ts) offer the opportunity
of observing high-energy neutrinos, through their interactions deep in ice
or water volumes where the detector modules are installed. The IceCube collaboration, in particular, has reported evidence for the existence of a high-energy lepton flux with increasing statistics over the years~\cite{Aartsen:2013jdh}. At present, the best-fit to the data looks quite compatible or even suggests an astrophysical interpretation of the observed signal, but no specific correlation with particular galactic or extragalactic sources has been identified, at least so far, thus the
origin of the IceCube events still remains a mistery. Many tentative hypotheses have been formulated to explain this signal~\cite{Fong:2014bsa}, taking into account that neutrinos produced in a number of astrophysical environments may actually reach us after travelling long distances undeflected by cosmic magnetic fields. Possibly interesting production sites range from galactic or extragalactic sources and their neighborhood, to dark matter (DM) populated regions where heavy DM decay or DM-DM annihilation could occur. In all cases, neutrinos produced by the interaction of cosmic rays (CRs) with the Earth's atmosphere represent indeed a background that one has to subtract
in order to disentangle the signal of truly astrophysical origin.
Actually, taking into account that the CR spectrum at the top of the
Earth's atmosphere is peaked in the region $E_{CR}$~$\sim$ $0.2~-~1$~GeV per nucleon for all ions, the bulk of atmospheric neutrinos, i.e., of neutrinos produced in the atmosphere, has energies $E_{\nu}$~$\sim$ $0.05~-~0.4$~GeV. These neutrinos are mainly the results of hadronic interactions
between the impinging CRs and the light atmospheric nuclei (mostly nitrogen
and oxigen), coming from intermediate charged pions and kaons, which subsequently decay
leptonically. However, at increasing CR primary energy, the decay lenghts of
these light mesons increase, up to and exceeding the transverse size of the atmosphere. In these conditions, $\pi^\pm$ and $K^\pm$ decay probabilities are suppressed with respect to those of their reinteraction, and $\nu$'s are mainly generated by other mechanisms, especially the atmospheric production of 
charmed mesons and baryons, followed by their prompt decay. Charmed hadrons in fact have decay lenghts smaller than $\pi^\pm$ and $K^\pm$: the prompt neutrino flux, initiated by charm, becomes larger than the conventional one, initiated by light mesons, at an energy called ``transition energy''. The exact value of this energy is still subject to sizable uncertainties, mainly reflecting those affecting highly energetic hadronic interactions in the atmosphere.
We have recently re-evaluated charm hadroproduction in the atmosphere~\cite{Garzelli:2015psa}, in the light of recent QCD theoretical and experimental progresses, triggered by the flourishing activities at hadron colliders, 
in particular the Large Hadron Collider (LHC): in this contribution we mainly review and further discuss the results presented more extensively in that paper.
  
\section{Charm hadroproduction in QCD and prompt neutrino fluxes}
\label{charmhadro}
\begin{figure}[ht]
\centering
\includegraphics[scale=0.54]{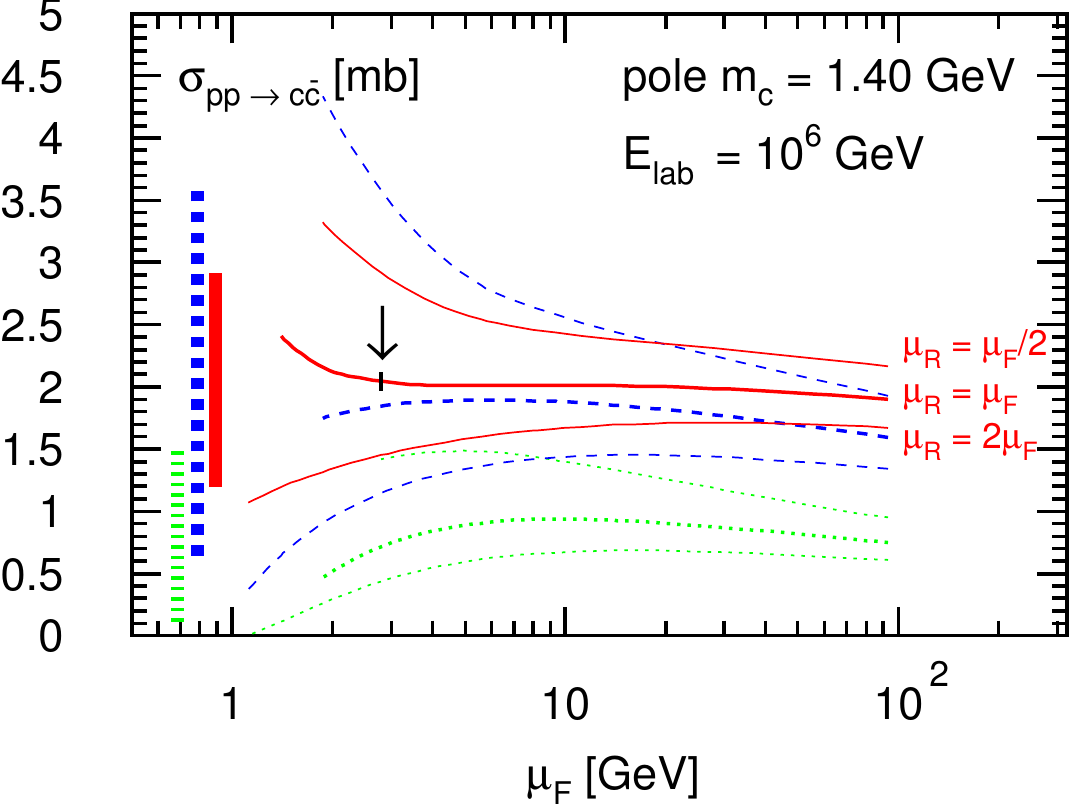}
\includegraphics[scale=0.54]{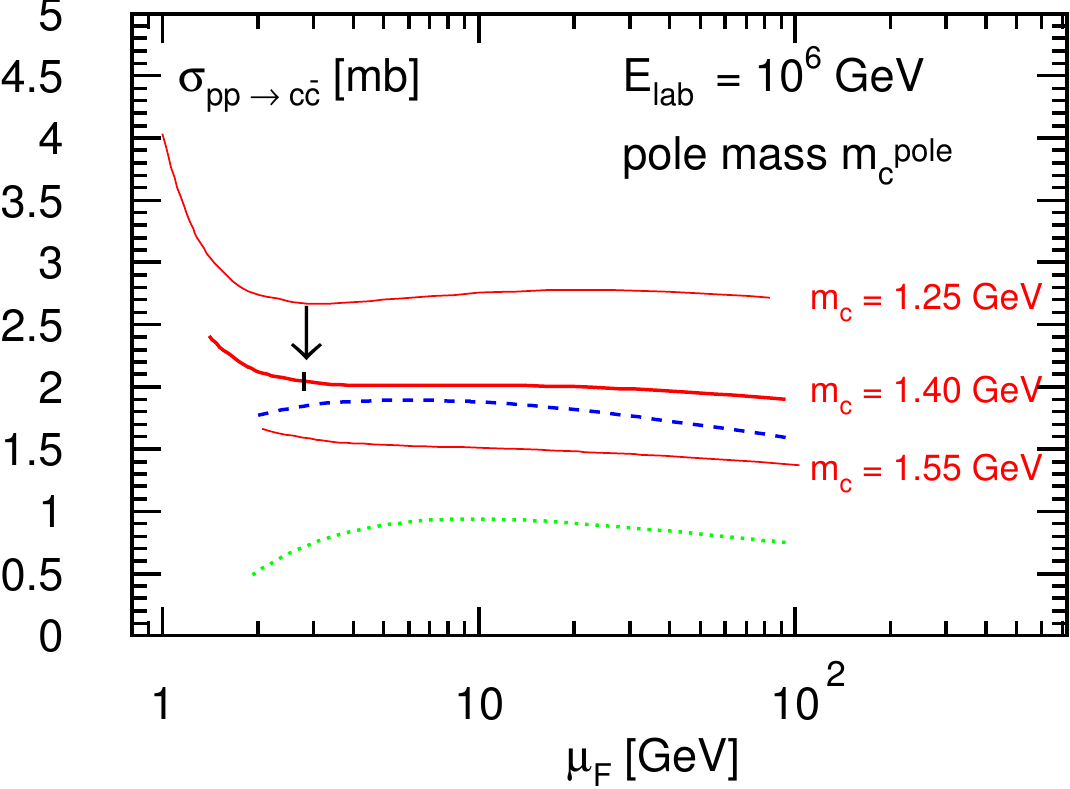}
\caption{Left panel: cross-section for $c\bar{c}$ hadroproduction as a function 
of $\mu_F$ for different values of $\mu_R$. Charm mass is fixed to $m_{charm}$ = 1.4 GeV. The predictions in red, blue and green refer to NNLO, NLO and LO approximations, respectively. At each fixed order, the central line refers to predictions with $\mu_R = \mu_F = m_{charm}$, whereas the upper and lower lines denotes predictions for $\mu_R = \mu_F/2$ and $\mu_R = 2 \mu_F$, respectively. 
Right panel: cross-section for $c\bar{c}$ hadroproduction as a function 
of $\mu_F$. $\mu_R$ is fixed to $\mu_R$ = $\mu_F$. Varying the charm mass value
in the range $m_{charm}$ = 1.4 $\pm$ 0.15 GeV gives rise to the three NNLO 
predictions denoted by red lines, whereas blue and green lines corresponds
to NLO and LO predictions for the central $m_{charm}$ value. }
\label{figmufmur}       
\end{figure}

In QCD the collinear factorization formalism allows to write the total
cross-section for charm hadroproduction as a convolution of a non-perturbative
component, involving the parton distribution functions (PDFs) and the
fragmentation functions (FF), with a perturbative part, given by the partonic
cross-sections for the production of charm quarks from initial state quarks
and gluons. In the following, we assume that the collision of a CR with an
atmospheric nucleus can be simply described as a superposition of
Nucleon-Nucleon ($NN$) interactions. Recent progress in pQCD gives nowaday the
possibility to calculate the total cross-section for $NN \rightarrow c\bar{c}$
pair hadroproduction including radiative corrections up to
next-to-next-to-leading order (NNLO), by properly combining NNLO PDFs and NNLO
partonic amplitudes. In practice, we obtained this cross-section as a function
of the $NN$ collision energy by extending the {\texttt{HATHOR}} event
generator~\cite{Aliev:2010zk}, originally developed for $t\bar{t}$
hadroproduction. The truncation of the QCD perturbative series at a fixed
order makes the theoretical predictions for the cross-sections dependent of
two unphysical scales: the renormalization scale $\mu_R$ and the factorization
scale $\mu_F$. Unfortunately, there is no unique unambiguous way derived from first principles to properly fix these scales. We fix $\mu_R$ and $\mu_F$ by looking for minimal sensitivity, i.e. 
in such a way to minimize the impact of radiative corrections when comparing next-to-leading order (NLO) to NNLO predictions. The point of minimal sensitivity approximately corresponds to the choice $\mu_F$~=~$\mu_R$~=~$\mu_0$~=~2~$m_{charm}$, denoted by an arrow in Fig.~\ref{figmufmur}a. Besides the total cross-section, we are interested in differential ones. Thus we translate this static scale into a dynamical one $\mu_0 = \sqrt{p_{T,charm}^2 + m_{charm}^2}$, to better catch the kinematical aspects of charm hadroproduction.  We take this as our central scale choice for the generation of differential distributions, and then we calculate scale uncertainties by considering the independent variation of $\mu_R$ and $\mu_F$ in the range [0.5 , 2] $\mu_0$, keeping out the two extreme combinations ($\mu_R$, $\mu_F$) = (2, 0.5) and (0.5,2) $\mu_0$, as prescribed in Ref.~\cite{Cacciari:2012ny}.

Furthermore, the $c\bar{c}$ hadroproduction cross-sections 
are sensitive to the precise value of charm mass, as shown in Fig.~\ref{figmufmur}.
Particular care has to be taken in the choice of this parameter especially when adopting the on-shell (OS) scheme to renormalize heavy-quark masses, because this scheme shows a poor perturbative convergence, actually worse with respect to
the $\overline{MS}$ scheme. We fix the charm mass value to $m_{charm}^{OS}$ = 1.40
$\pm$ 0.15 GeV. This way the cross-section in the OS scheme approximately
reproduces that in the $\overline{MS}$ scheme. We notice that the accuracy on
the pole mass is also limited by the renormalon ambiguity. 
 
Further QCD uncertainties are related to the PDF choice. At present, many uncertainties still exist at low Bjorken $x$ ($x$ $<$ 10$^{-4}$), due to lack of data. However this region is progressively important when going to higher collision energies, which is a crucial issue for astrophysics because CRs in the tail of the CR energy spectrum are even more energetic than the beams at colliders.  Recently, LHCb data on charm and bottom hadroproduction in mid-pheripheral collisions have been used to perform PDF fits down to lower values of $x$ ($x$ $\sim$ 10$^{-6}$). The first fit in this direction was proposed by the PROSA collaboration in Ref.~\cite{Zenaiev:2015rfa}, followed by Ref.~\cite{Gauld:2015yia} applying the same concept to the NNPDF PDFs. In this paper we use the ABM11 PDFs~\cite{Alekhin:2012ig}. We notice that, although these PDFs do not take into account any LHCb data, their extrapolation in the range 10$^{-6}$ $<$ x $<$ 10$^{-4}$ turns out to lie within the PROSA PDF uncertainty band. We evaluate PDF uncertainties by considering the 28 ABM11 variations, accompanying the central fit available in the LHAPDF 6.1.5 interface. 

\begin{figure}[ht]
\centering
\includegraphics[scale=0.22]{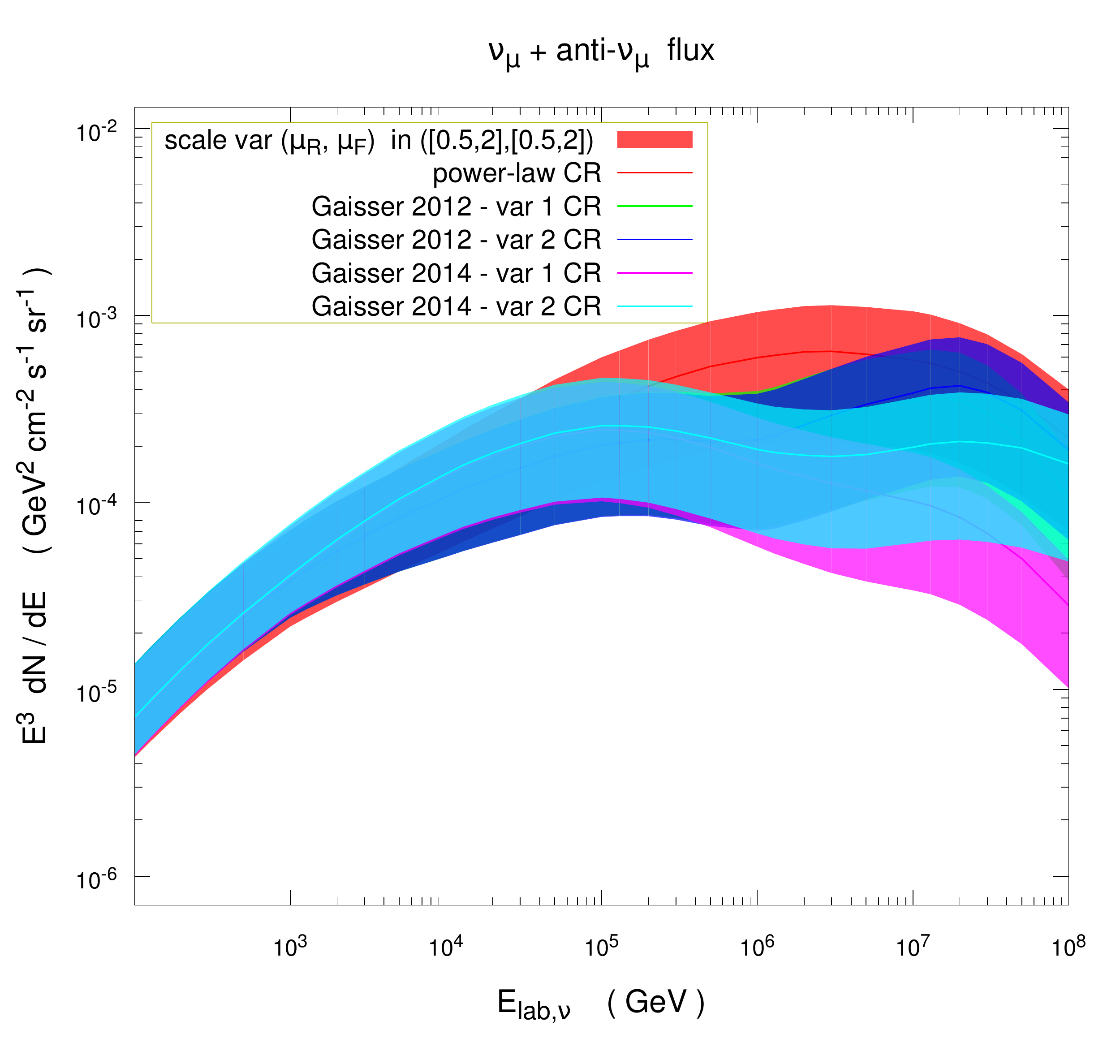}
\includegraphics[scale=0.22]{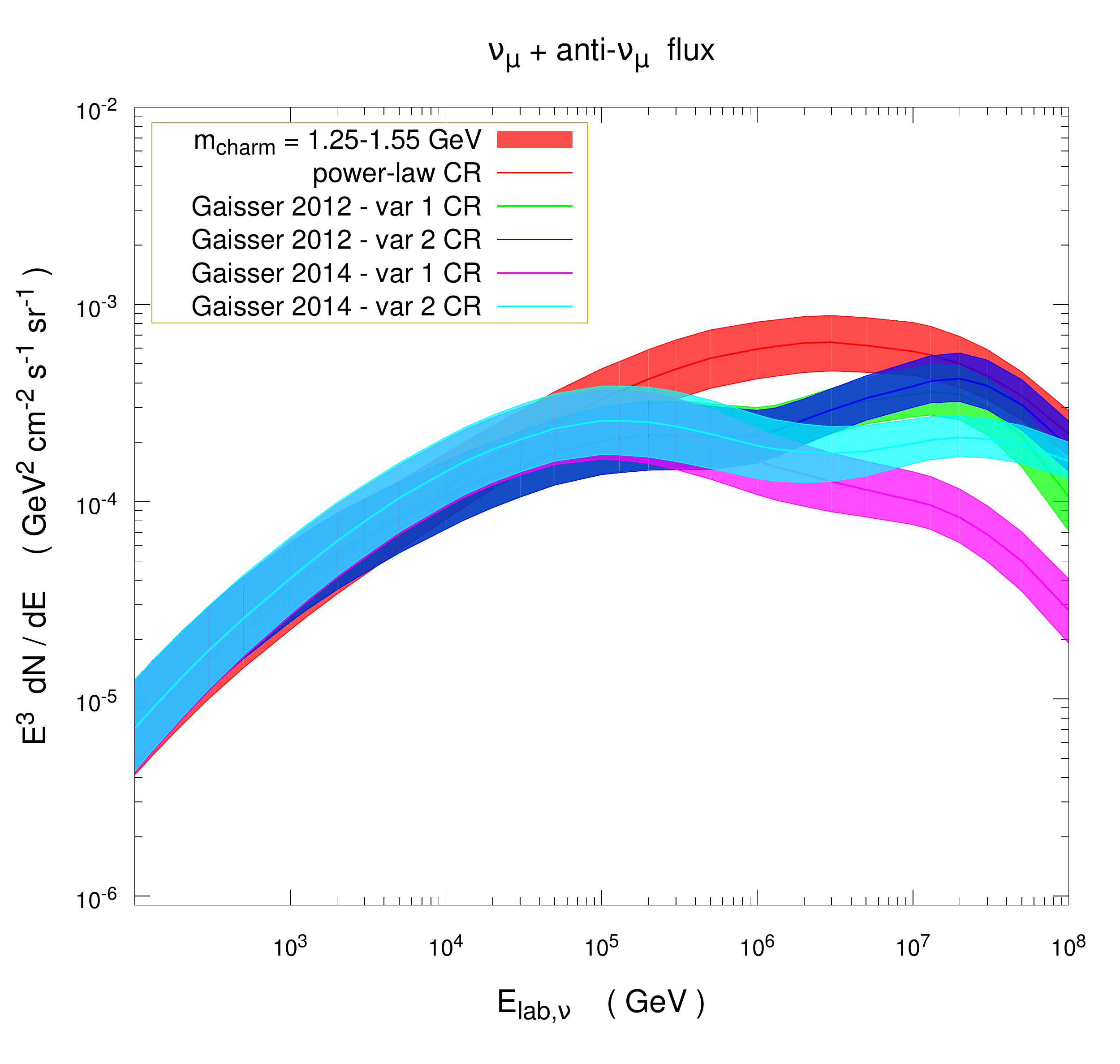}\\
\includegraphics[scale=0.22]{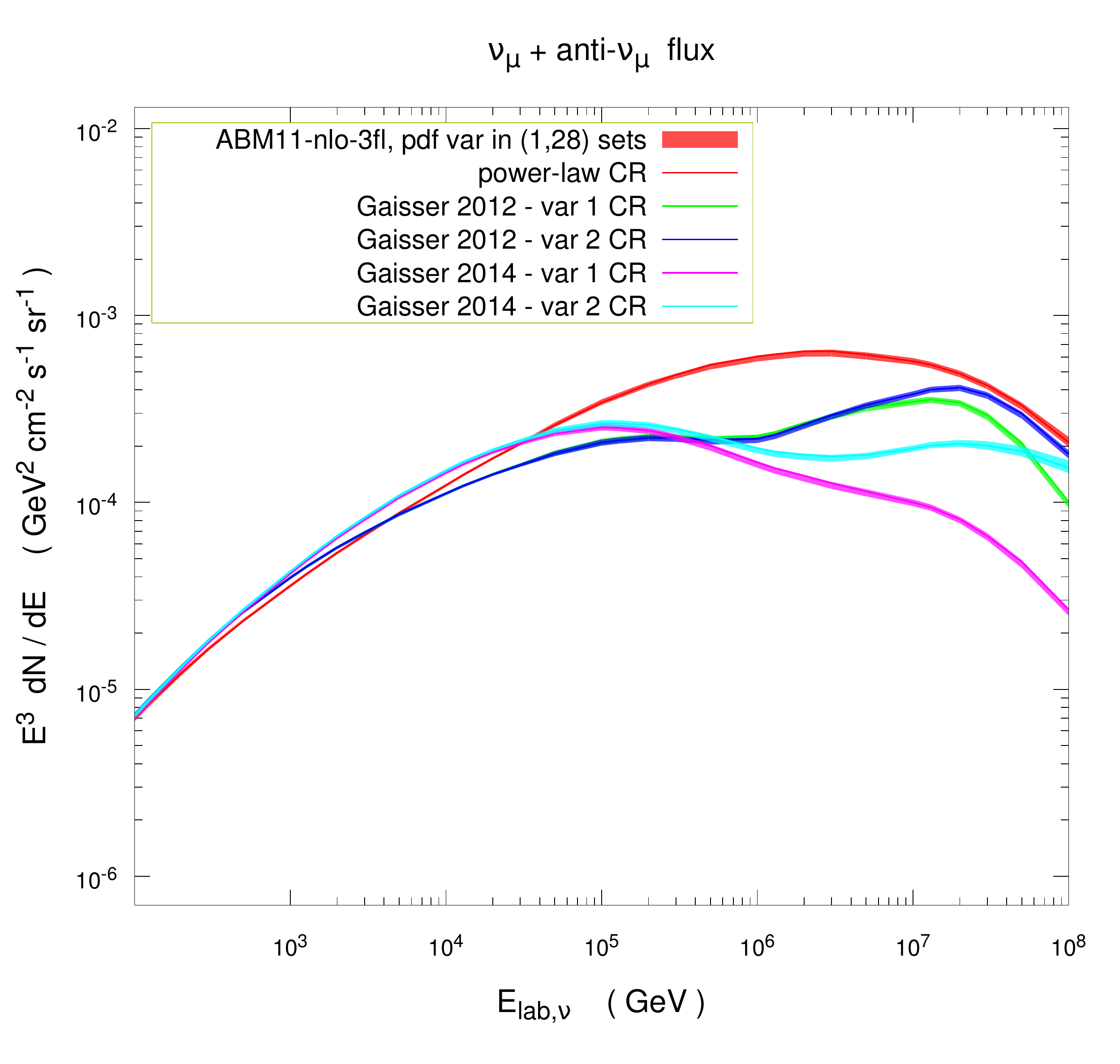}
\includegraphics[scale=0.22]{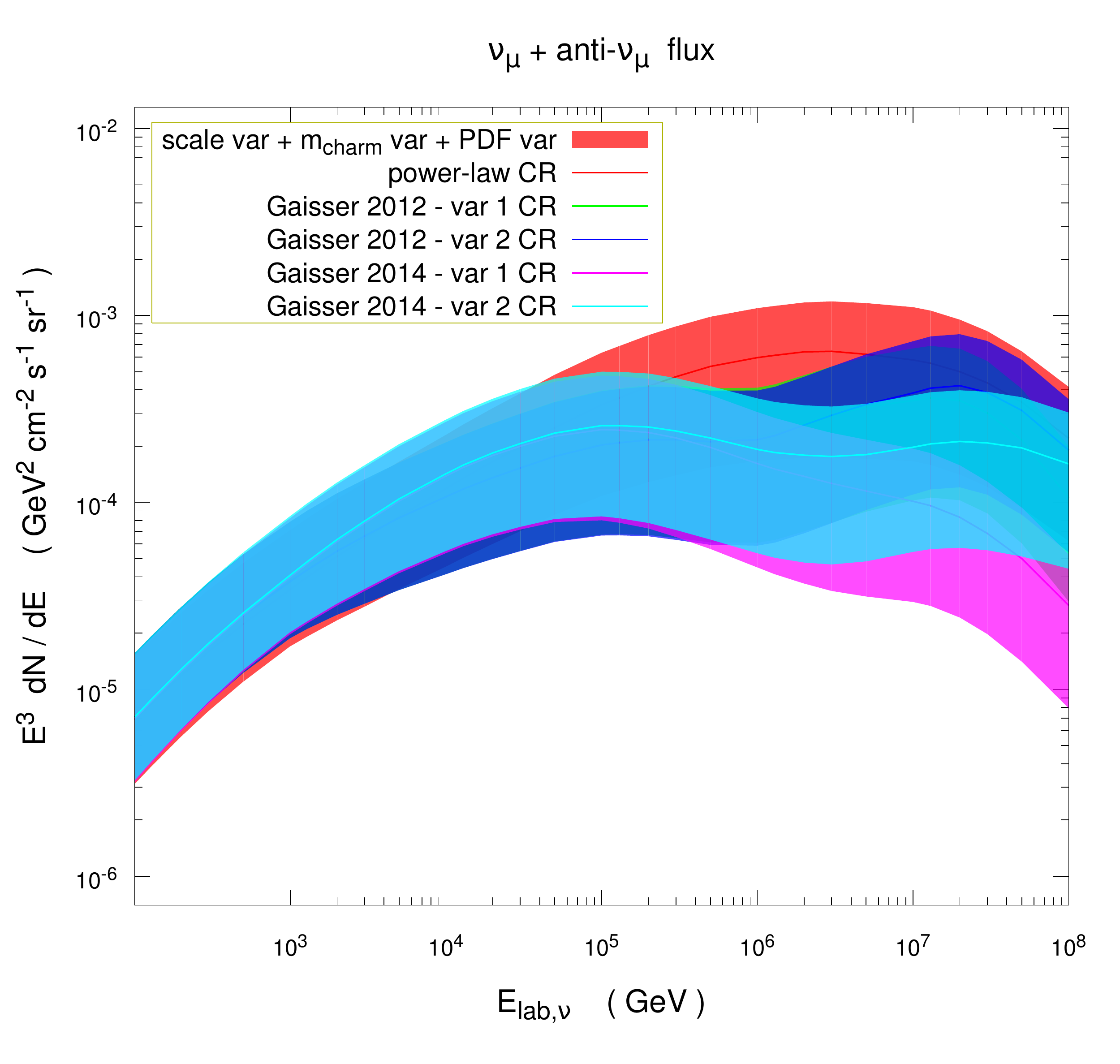}

\caption{Prompt ($\nu_\mu$ + $\bar{\nu}_\mu$) fluxes as a function of neutrino energies: QCD uncertainties due to scale, mass, PDF variation and their combination in quadrature are shown in panel a, b, c, d, respectively. For each panel, predictions related to the use of different primary CR all-nucleon spectra, are shown by different colors: the broken power-law spectrum is in red, whereas four more modern spectra (see Ref.~\cite{Gaisser:2013bla} and references therein) are shown with other colors.}
\label{fig2}
\end{figure}

Neutrino fluxes were computed by solving a system of coupled differential equations regulating particle evolution in the atmosphere, by using the $Z$-moment approach~\cite{Lipari:1993hd}. Input of the $Z$-moments are differential cross-sections for charmed hadron hadroproduction. Unfortunately, these distributions are not yet available at NNLO. We thus computed them through a NLO QCD + parton shower + hadronization approach, as provided by {\texttt{POWHEGBOX}}~\cite{Alioli:2010xd} + {\texttt{PYTHIA}}~\cite{Sjostrand:2006za}. Charm mass, PDF and scale input of the computation and their intervals of variation were fixed on the basis of NNLO information, as explained above.  
The effect of uncertainties due to scale, charm mass and PDF variation on
($\nu_\mu$~+~$\bar{\nu}_\mu$) prompt fluxes is shown in Fig.~\ref{fig2}a, b,
c, respectively. In the last panel, Fig.~\ref{fig2}d, the combination of those
QCD effects, summed in quadrature, is also shown. The total contribution of
additional uncertainties related to hadronization and multiple particle
interactions amounts to several ten percent. We observe that, among all QCD
uncertainties, the one due to scale variation dominates. We also notice that at energies above the PeV scale, the contribution of astrophysical uncertainties, related to the unknown composition of the CR spectrum at primary energies above the knee, becomes progressively larger and ends up in dominating over the contribution from QCD uncertainties at ultra-high neutrino energies. 
The region around and above the PeV will be better probed by upgraded VLV$\nu$Ts, like the IceCube-Gen2 project~\cite{Aartsen:2014njl}, providing an increase of the instrumented volume from $\sim$ 1 km$^3$ to $\sim$ 10 km$^3$. By the time they will start data taking, it would be advisable to reduce all uncertainties (both those from QCD and those from astrophysics) affecting the computation of prompt neutrino fluxes at high energies. Our most up-to-date predictions are available at {\texttt{http://promptfluxes.desy.de}}.

\bibliography{biblio.bib}

\end{document}